# Superheated Microdrops as Cold Dark Matter Detectors

**J. I. Collar**

Department of Physics and Astronomy
University of South Carolina
Columbia, South Carolina 29208, USA

and

Groupe de Physique des Solides (UA 17 CNRS)
Universites Denis Diderot (Paris 7) and P. & M. Curie (Paris 6)
2, Place Jussieu, 75251 Paris Cedex 05, FRANCE

electronic address: ji.collar@sc.edu

astro-ph/9607150  29 Jul 1996

**Abstract:**
**It is shown that under realistic background considerations, an improvement in Cold Dark Matter sensitivity of several orders of magnitude is expected from a detector based on superheated liquid droplets. Such devices are totally insensitive to minimum ionizing radiation while responsive to nuclear recoils of energies ~ few keV. They operate on the same principle as the bubble chamber, but offer unattended, continuous, and safe operation at room temperature and atmospheric pressure.**

A number of experimental efforts aim at the detection of nuclear recoils produced by the elastic scattering of Weakly Interacting Massive Particles (WIMPs) off target nuclei [1]. The next generation of Cold Dark Matter (CDM) detectors will require a sensitivity $\lesssim$ 1 recoil / kg / day to discover or rule out the neutralino, a CDM candidate arising from supersymmetric extensions of the Standard Model [2]. To achieve this, future devices must have the ability to distinguish nuclear recoils in the ~ keV energy range from similar energy depositions by minimum ionizing radiations, still present in ultra-low background underground detectors [3]. Several schemes have been proposed in this respect, such as the simultaneous detection of ionization and phonons [4]; their realisation has proven to be a non-trivial task. In addition to this, large detector masses are necessary for unambiguous WIMP identification through the small modulations characteristic of the WIMP recoil signal [5-7]: a target mass $\gtrsim$ 100 kg is needed to detect the ~ 5 % yearly change in recoil rate of ref. [7] in a modest period of time (few years). Planned experiments are far from simultaneously meeting these demands.

Ideally, a WIMP detector should be sensitive only to recoils like those produced by fast neutrons. One such detector exists; superheated liquid droplet neutron detectors, also known as "bubble detectors" (BDs) and first proposed by Apfel [8], are strictly sensitive to high-Linear Energy Transfer (LET) radiation (the unrestricted LET is the amount of energy dissipated by a radiation per unit path length, $dE/dx$). Energetic muons, gamma rays, X-rays and beta particles have a LET well below the activation threshold of BDs, which is typically $\gtrsim$ 200 keV / µm at room temperature. Bubble detectors are totally insensitive to gamma rays of energy < 6 MeV at operating temperatures T < 30° C [8-10]. This and several other advantages can make of them an optimal device for neutralino matter searches.

A bubble detector consists of a dispersion of droplets (radius 25-75 µm) of a superheated liquid, fixed in a viscous polymer or aqueous gel [8,11]. Several techniques exist for the production of BDs [12,13]. The metastable superheated state is maintained indefinitely at room temperature and 1 atm. Mechanical energy is stored in each droplet, which behaves like a miniature bubble chamber. The energy deposition of a particle can release this energy, triggering the vaporization of the droplet and forming a visible gas bubble (diameter ~ 1 mm). Depending on the viscoelasticity of the medium, the bubbles remain fixed in it, providing a visual record of the radiation dose and simple optical reading, or rise to collect above it, where the volume of evolved gas can be measured. Alternatively, the sound released by the sudden vaporization can be registered by a piezoelectric transducer [12,14]. The performance of BDs has been measured over a period of four years, without a significant loss of sensitivity [9]. Polymer-based BDs do not display shock-induced bubbles when dropped from heights of several meters [15]. The difficulties inherent to the handling of large volumes of superheated liquid are absent in BDs; practically, conventional bubble chambers used in nuclear experiments are "dirty" in the sense that there is homogeneous nucleation caused by large scale statistical density fluctuations in the bulk of the liquid and heterogeneous nucleation from contact with the container walls, gaskets, etc. This boiling tends to recompress the liquid and the bubble chamber needs to be pressure-cycled every few seconds. The low event rate necessary for a competitive dark matter experiment seems difficult to obtain in bubble chamber proposals [16]. Safety concerns such as explosive boiling of a large volume of superheated liquid (contact vapour explosion) disappear; the vaporization of a single droplet does not generate avalanche effects and BDs operate in a passive, unattended fashion, with no external power supply. The probability of spontaneous vaporization of droplets is extremely small; a droplet of radius as large as ~ 1 mm is stable against homogeneous nucleation for > $10^6$ y at T as high as 50°C [13,17]. The rate of spontaneous bubble formation in commercial BD personal dosimeters is accounted for by the cosmic neutron flux [9,18]. The volume filling factor of the superheated liquid is kept low (~1 %) in commercial BDs to avoid an excessive response to this background.

The mechanism of bubble nucleation under irradiation has been studied by several authors [10,12,17,19,20] and is based on the thermal spike model of Seitz [21]; an intense energy deposition along a particle's path can provide enough localized heating to create bubbles of a critical size or larger. If a vapour bubble grows larger than a critical radius $r_c(T)$ (~ few tens of nm [19]), it becomes thermodynamically unstable and continues to expand evaporating all of the droplet's liquid. The conditions necessary for radiation induced nucleation are two [17,19]:

(1) The total energy deposited must be larger than a critical energy for bubble formation, $E_c(T)$, computed from the sum of the thermodynamically reversible processes of vaporization, formation of bubble surface and bubble expansion against the gel, and

(2) the stopping power of the particle in the target material must be such that $E_c(T)$ is deposited within a small distance L (T) of order $r_c$:

$$(dE/dx)\, L(T) \geq E_c(T). \qquad (1)$$

This second condition is responsible for the insensitivity of BDs to low-LET radiations. Note that the response of BDs does not depend on droplet size, as long as they are not smaller than $r_c$ and large enough to produce visible bubbles if optical reading is used.

Several liquids have been tested in BDs [10]. Freon®-12 ($CCl_2F_2$) [22] is by far the best for a CDM search due to its very low critical energy; e.g., $E_c(30°C) \approx 5$ keV, $E_c(20°C) \approx 16$ keV, increasing to $E_c(0°C) \approx 200$ keV [19]. The response of Freon-12 BDs to thermal, fission, and monochromatic neutrons has been investigated and is in good agreement with theoretical models [10,12,17,19]. Freon-12 is the main concern of this paper.

The value of L(T) for Freon-12 has been measured using a $^{252}Cf$ neutron source [17,19]. It varies in the range 1.0 $r_c$ - 1.5 $r_c$ for T between 0° C - 35° C. Figure 1 displays the requirements for bubble nucleation in Freon-12 at different T. The total stopping power of alpha particles and recoiling atoms in the liquid, obtained from the code TRIM92 [23], is also shown. While a Freon-12 BD is sensitive to Cl recoils down to $E_c$ in the temperature range envisioned for a CDM search (~20° C - 30° C), this is not the case for F or C recoils. For spin-independent coherent WIMP interactions, where the scattering cross section varies with the square of the number of nucleons in the target nucleus, this feature does not reduce the expected WIMP bubble production rate by much, Cl being the heaviest of all Freon-12 components. However, one advantage of this detector resides in its high fluorine content; this nucleus has the largest expected interaction rate for axial-vector (spin-dependent) neutralino scattering [24]. Operating temperatures close to 30° C must be used in order to maximize the sensitivity to this sector of the neutralino parameter space. Above ~ 27° C the radiopurity of the gel becomes a concern [18], due to the possibility that alpha particles from U and Th contaminants produce bubble nucleation. The absence of alpha-induced nucleations below 25 - 30° C has been recently demonstrated using actinide-doped BDs [25].

Several types of neutron interactions meet both requirements for bubble nucleation [10]: (1) elastic scattering, (2) inelastic scattering, (3) $^{35}Cl(n,p)^{35}S$, and (4) $^{35}Cl(n,\alpha)^{32}P$. The last two are predominant for thermal neutrons. A simple model has been developed to predict the response of BDs to neutrons [10,12,17,20]; this energy-dependent response function, $P(E_n)$, is calculated as:

$$P(E_n) = \psi(E_n) V \sum_i N_i \sum_j \sigma_{ij}(E_n) S_{ij}(E_n, T), \qquad (2)$$

where $E_n$ is the incident neutron energy, $\psi(E_n)$ is the neutron fluence, V is the total volume of superheated liquid, $N_i$ is the number of atoms per unit volume of the ith atomic species in the liquid, $\sigma_{ij}(E_n)$ is the neutron cross section of the jth interaction with the ith atomic species, and $S_{ij}(E_n,T)$ is the "superheat factor", i.e., the fraction of the recoil nuclei with kinetic energy above the minimum (threshold) energy, $E_{thr}(T)$, that satisfies both requirements for bubble nucleation. Calibrations using monochromatic neutrons are in excellent agreement with this model's predictions [10]. The Freon-12 response function has an advantageous feature for a CDM search: it drastically drops from an approximately constant value of $P(E_n) \sim 10^{-1}$ (in units of bubbles / neutron per cm$^2$ / cm$^3$ of Freon-12) for $E_n \gtrsim 200$ keV, to $P(E_n) \sim 10^{-6}$ for $E_n \sim$ few tens of keV, slowly increasing to $P(E_n) \sim 10^{-3}$-$10^{-4}$ for thermal $E_n$ [10]. This allows for very efficient shielding of the underground neutron spectrum by simply surrounding the detector with water, which can then be used to simultaneously regulate the operating temperature (as in a double-bath). Fast neutrons, to which Freon-12 is most sensitive, fall after moderation in the region of diminished $P(E_n)$.

The predictability of the response of BDs to neutron recoils makes one confident that equally reliable predictions can be generated for WIMP-induced recoils. The temperature dependence of the threshold energy $E_{thr}$ allows for the measurement of a differential recoil energy spectrum by running the detector at different temperatures. This differential rate depends strongly on the mass and coupling of the scattered particle and can be used to differentiate a true WIMP signal from the backgrounds discussed below.

Elastic scattering is by far the most important mode of WIMP interaction and Eq. (2) is thereby simplified by the removal of the jth index. The recoil energy, $E_{rec}$, transferred to the ith atomic species is determined from the differential cross section, $\frac{d\sigma}{dE_{rec}}$, of the particular WIMP candidate under consideration. A heavy

"neutrino", i.e., a generic massive neutral particle is used in this calculation, having the differential cross section for elastic scattering [26]:

$$\frac{d\sigma}{dE_{rec}} \propto G\,N^2 \frac{m_R^2}{E_{rec}^{max}(m_\chi)} F(q^2), \qquad (3)$$

where G is a coupling constant, $m_R$ is the reduced mass of the WIMP-nucleus system, $E_{rec}^{max}(m_\chi)$ is the maximum recoil energy for a given WIMP mass $m_\chi$, and N is the number of neutrons in the target nucleus. The term $F(q^2)$ is a form factor accounting for the loss of coherence for very massive WIMPs. The exponential approximation in ref. [3],

$$F(q^2) = \exp\left(\frac{-2 \cdot M \cdot E_{rec} \cdot \varepsilon^2}{3 \cdot h^2}\right), \qquad (4)$$

is adequate even for the heaviest nuclei [27] ($\varepsilon$ is the nuclear radius and M is the mass of the recoiling nucleus). The choice of a heavy "neutrino" as the WIMP is justified by the simplicity in the computation of its rate of interaction and energy transfer to the nucleus (neutralino cross sections are far more parameter-intensive). Moreover, the scalar neutralino coupling from Higgs boson exchange [28] ($\sigma \sim m_R^2(N+Z)^2$, Z is the atomic number) prevails over spin-dependent channels for most neutralinos with a Zino-Higgsino mixture [29,30]; their differential scattering cross-section depends on $E_{rec}$ only through the same form factor of Eq. (4) [27,30]. In this simple case, $F(q^2)$ alone defines the energy distribution of recoiling nuclei. The collision kinematics for these neutralinos and heavy "neutrinos" are then the same to a good approximation, and the rates of bubble nucleation presented below apply to this important neutralino sector after scaling by the coupling constant in Eq. (3).

The rate of interaction at a given recoil energy is computed by weighted integration of $\frac{d\sigma}{dE_{rec}}$ as is customary in WIMP direct searches [3]. The relevant halo parameters used here are $v_{Earth}$ = 260 km / s (annual average of Earth's speed through the halo), $v_{dis}$ = 270 km / s (dispersion in the galactic CDM speeds), $v_{esc}$ = 550 km / s (local galactic escape velocity) and $\rho_{halo} \approx 0.4$ GeV / $c^2$ / $cm^3$ (local halo density) [3]. The superheat factor for WIMP recoils, $S_i$, is equal to the recoil rate between $E_{thr}$ and $E_{rec}^{max}(m_\chi)$, divided by the total recoil rate. This $S_i$ is shown in fig. 2 for different T; note that $E_{thr}$ for carbon is so large for $T \lesssim 25°$ C (fig. 1) that few nucleations from carbon recoils are expected. The large change in $S_F$ from T= 25° C to 30°C is due to the very different $E_{thr}$ values for F (~110 keV and 5 keV, respectively). The equivalent of Eq. (2) can be employed now to obtain the rate of WIMP bubble nucleation, using the averaged values of WIMP fluence and cross section over their velocity distribution in the Earth's frame [3]. A slightly more accurate method is to integrate the recoil rate from $E_{thr}$ to $E_{rec}^{max}$ for each atomic species, weighting the results by the species' abundance and summing them. The so obtained response function is plotted in fig. 3 for a heavy Dirac neutrino. At T= 25° C, the seasonal change in bubble production from the yearly modulation of ref. [7] is ~ 5% for $m_\chi \gtrsim 100$ GeV / $c^2$ and larger for lighter $m_\chi$ (~ 30% for $m_\chi$ = 10 GeV / $c^2$) [18].

Perhaps the simplest implementation of BDs as WIMP detectors will consist of modular containers of dimension ~1 m x 1 m x 0.1 m, filled with BD, and immersed in a temperature-controlled water double-bath in an underground laboratory. Such a volume contains, even at the low commercial filling factor, a target mass ~ 1 kg. With large enough BD masses the technique can be extrapolated to the detection of solar and stellar collapse neutrinos [18], taking advantage of the coherent enhancement to their neutral-current scattering cross section [31].

Even in such an environment, certain sources of high-LET background radiation will be unavoidable. A typical [32] underground neutron flux has been measured in the Gran Sasso laboratory (depth ~ 3800 meters of water equivalent, m.w.e.) [33]. This flux is $\Phi_n$ ~ $3.8 \cdot 10^{-6}$ $cm^{-2}$ $s^{-1}$ for $10^{-3}$ eV $< E_n <$ 25 MeV. This is expected from the typical neutron production rate in rocks, $N_n$ ~ $2.5 \cdot 10^{-7}$ neutrons $g^{-1}$ $s^{-1}$ [34] and the attenuation length for fast neutrons in granite, $\lambda_{granite}$ = 30 g / $cm^2$ [35]; the flux inside a (thick) 4π "shield" (rock walls) is approximately given by:

$$\Phi_n \approx N_n \lambda \sim 7.5 \cdot 10^{-6} cm^{-2} s^{-1}, \qquad (5)$$

in rough agreement with the observed value. Folding the measured differential $\Phi_n$ of ref. [33] with the response $P(E_n)$ in ref. [10], one obtains the expected rate of bubble production by the unmoderated neutron flux in a typical underground laboratory (fig. 3). The water-moderated flux, $\Phi'_n$, is calculated by means of the mass removal cross section, i.e., $\lambda^{-1}$ [35]:

$$\Phi'_n = \Phi_n\, e^{-t/\lambda} \qquad (6)$$

where t is the distance travelled in water and $\lambda_{H_2O} = 10.1\, g/cm^2$ [35]. Strictly speaking, Eq. (6) applies only to the fast component of the neutron flux (~ $0.7 \cdot 10^{-6}$ cm$^{-2}$ s$^{-1}$ for $E_n \gtrsim 200$ keV [33], i.e., in the energy region of maximum $P(E_n)$); the moderated neutrons can still produce nucleations, albeit with a largely diminished $P(E_n)$. Here a conservative $P(E_n) = 10^{-4}$ (see above) is assumed for those. The obtained expected rate of bubble nucleation is shown in fig. 3 for a modest $t \sim 70$ g/cm$^2$. A more precise Monte Carlo simulation of the moderated neutron spectrum, able to determine the optimal value of t, is underway [18]. Another concern is the neutron flux from U and Th impurities in the water shielding itself; the contribution from ($\alpha$,n) reactions is ~ 5 times larger than that from spontaneous fission of $^{238}$U [34,36]. Concentrations of U and Th as low as ~ 0.01 ppb are common even in tap water [37]. This gives an equilibrium alpha emission of ~ 300 kg$^{-1}$ day$^{-1}$. Taking a representative ($\alpha$,n) yield of ~ 1 neutron per $10^6$ alphas [38], a production rate of $N_n \sim 3.5 \cdot 10^{-12}$ neutrons g$^{-1}$ s$^{-1}$ is obtained. Use of Eq. (5) for a 4$\pi$ water shield gives a neutron flux at the BD of $\Phi_n \sim 3.5 \cdot 10^{-11}$ cm$^{-2}$ s$^{-1}$ from water radioimpurities. Allowing the maximum response to these neutrons, $P(E_n) \sim 10^{-1}$, results in $\sim 2 \cdot 10^{-4}$ bubbles / kg of Freon-12 / day (fig. 3). Finally, the neutron production by muons in rock is $N_n \sim 8 \cdot 10^{-12}$ neutrons g$^{-1}$ s$^{-1}$ at 3200 m.w.e. [34]; from the systematics of the underground rate of production of neutrons via ($\mu$,n) in different materials [39], this production rate should be down by a factor ~ 4 in water, resulting in $\sim 1 \cdot 10^{-4}$ bubbles / kg of Freon-12 / day from this channel (fig. 3). According to the classical theory of homogeneous nucleation [40], a temperature T > 90 °C would be needed to provoke this same rate spontaneously in superheated Freon-12 at 1 atm.

Two other sources of background, internal to Freon-12, must be contemplated. First, a recoiling daughter from alpha emission in U and Th impurities is able to produce nucleation. For instance, a $^{206}$Pb alpha-recoil has an energy ~ 103 keV, range ~ 0.08 μm and dE/dx ~ 1.8 MeV / μm in liquid Freon-12 [23]. An alpha-decay rate of ~ $3 \cdot 10^4$ kg$^{-1}$ day$^{-1}$ is expected for an equilibrium concentration of U and Th of 1 ppb. Fortunately, radioimpurities in cryogenic liquids can be frozen-out to levels < $10^{-15}$ g/g [41]; two alpha decays have been registered in a 6 kg liquid Xe detector after ~ 100 days [42]. A similar radiopurity in Freon-12 yields a nucleation rate comparable to that from the moderated neutron flux of fig. 3. At this level of radiopurity, a contribution from fission fragments is entirely negligible. Finally, Auger electrons from $^{36}$S are high-LET radiations, able to produce bubble nucleation; $^{36}$Cl ($T_{1/2} = 3 \cdot 10^5$ y) is expected to be present in small concentrations in Freon-12. The K-shell binding energy in S (2.5 keV) is an upper bound to the energy deposited by this process. This is below $E_c(30°C)$ and does not interfere with this search. Second order processes such as fission fragments originating in the gel and alpha reaction products will be treated elsewhere [18].

An expected WIMP exclusion plot can be generated by taking the largest of all the above background estimates (,i.e., moderated neutrons) as the signal detected. This is shown in fig. 4, together with present exclusions from underground Ge detectors [3]. Bubble detectors will be able to explore, in principle, a vast region of the neutralino parameter space.

In conclusion, BDs are mature enough to offer an excellent alternative for WIMP direct detection. The simplicity and low cost of these detectors, together with their inherent background rejection and the possibility of using large target masses, promise a great leap in CDM sensitivity. The development of a dedicated bubble detector has started in the framework of the Paris-Zaragoza-Lisbon-South Carolina collaboration.

Acknowledgements:
I have profited from conversations with F.T. Avignone, T. Girard, A. Morales, J. Morales and G. Waysand. I thank B. Sur for calling my attention to recent work by Zacek [16] before the completion of this work. I am indebted to the Fondation Robert Schuman for support during the 1994-1995 academic year and to the Groupe de Physique des Solides for their hospitality during the completion of this work.

Fig. 1

Linear Energy Transfer of recoil atoms and alpha particles in liquid Freon-12 ($\rho = 1.3$ g/cm$^3$), as extracted from TRIM92 [23]. Horizontal and vertical bars mark the minimum LET and critical energy, $E_c$, respectively, necessary for bubble nucleation (see text). At a given temperature, only recoils falling in the upper right quadrant inscribed by the corresponding lines can produce droplet vaporization; the intersection of the curves and the boundaries of the quadrants defines the threshold energy, $E_{thr}(T)$, for each atomic species.

Fig. 2

Superheat factor for recoils in Freon-12 (dotted lines = F, dash-dot = C, solid = Cl) from WIMP scattering via a scalar (spin-independent) coupling, as a function of the WIMP mass. These curves are common for heavy neutrinos and neutralinos with a Zino-Higgsino mixture. The operating temperature of the bubble detector is indicated for each curve.

Fig. 3

Bubble nucleation rate per kg of Freon-12 produced by a heavy Dirac neutrino composing the galactic halo ($\rho_{halo} \approx 0.4$ GeV/c$^2$/cm$^3$). The response to other particles with predominant spin-independent couplings is scaled down by their coupling constant, G (Eq. (3)). The characteristic variation of the response with operating temperature allows for the identification of the WIMP mass, $m_\chi$. The horizontal lines mark the nucleation rates expected from high-LET backgrounds existing in an underground laboratory (see text). For comparison purposes, the sea-level response to cosmic neutrons is $\sim 8 \cdot 10^3$ bubbles / kg / day. Bubbles produced by $\alpha$-particles emitted in the gel are not included (they produce nucleation only at T $\gtrsim$ 27° C [25]).

Fig. 4

Expected exclusion plot for WIMPs with scalar couplings from a Freon-12 BD after ~ 1 kg-y of data acquisition (solid lines, T as indicated). The water-moderated underground neutron spectrum of fig. 3 is conservatively assumed to be the predominant background. Freon cross sections are normalized to Ge for comparison purposes. Also shown: a) present exclusion limits from underground Ge experiments (UZ/PNL/USC collaboration [3]), b) expected improvement that could be obtained with a similar Ge detector able to reject 99 % of the low-LET background, c) cross section for a heavy Dirac neutrino (coherence loss included), and (shaded) regions populated by some neutralino candidates; d) minimal SUSY, e) GUT [43].

fig. 1

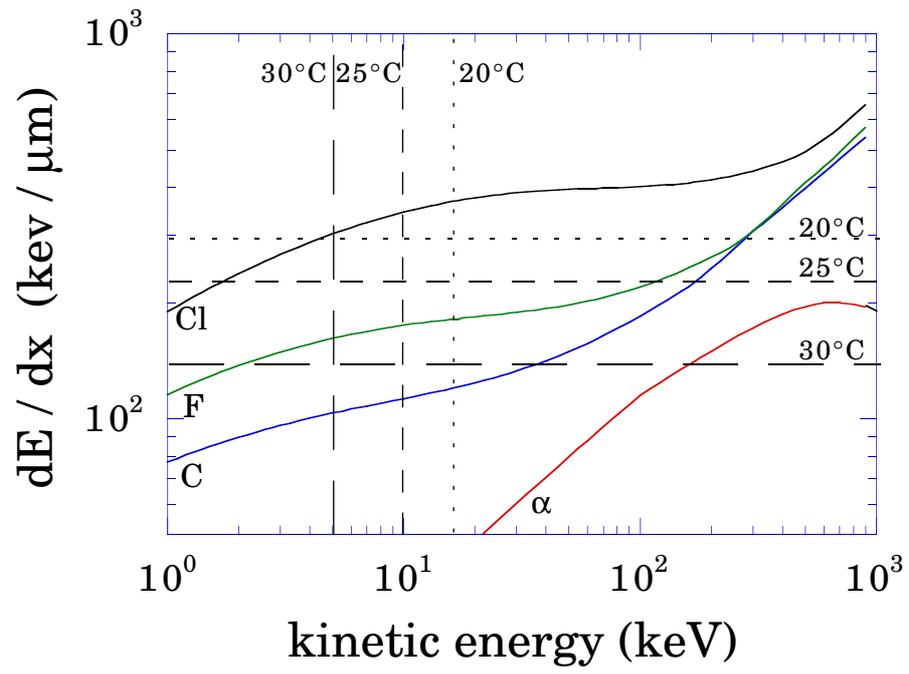

fig. 2

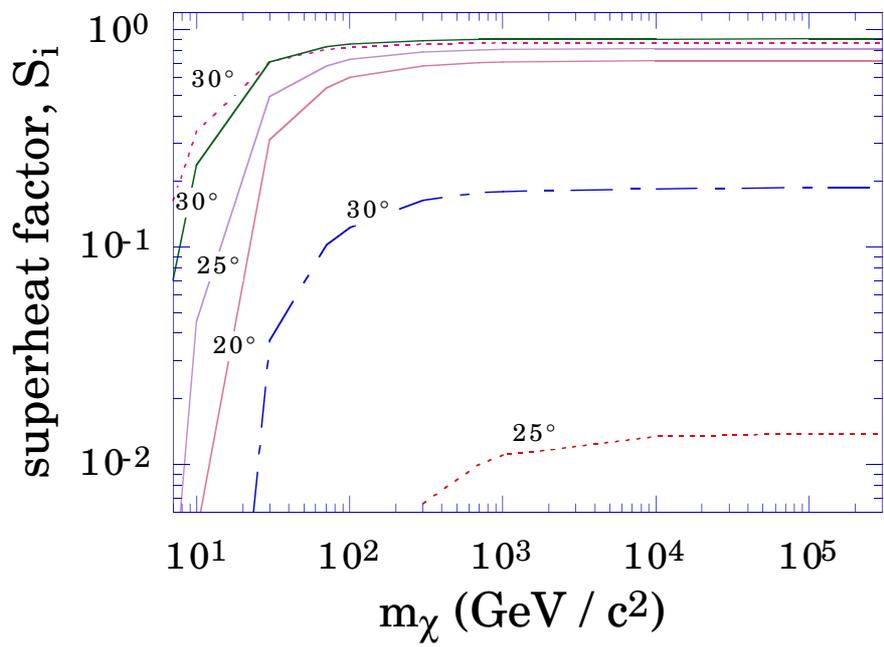

fig. 3

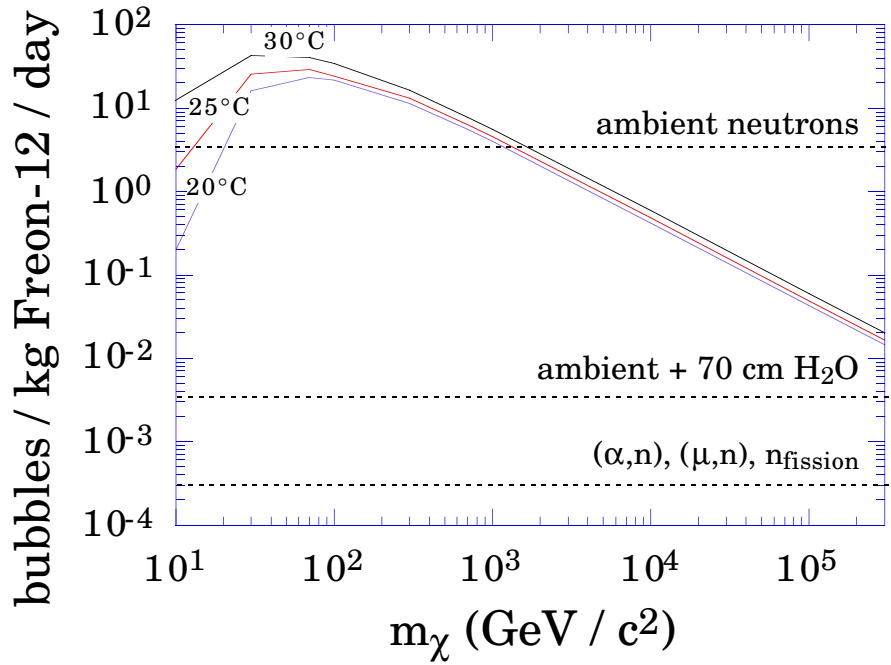

fig. 4

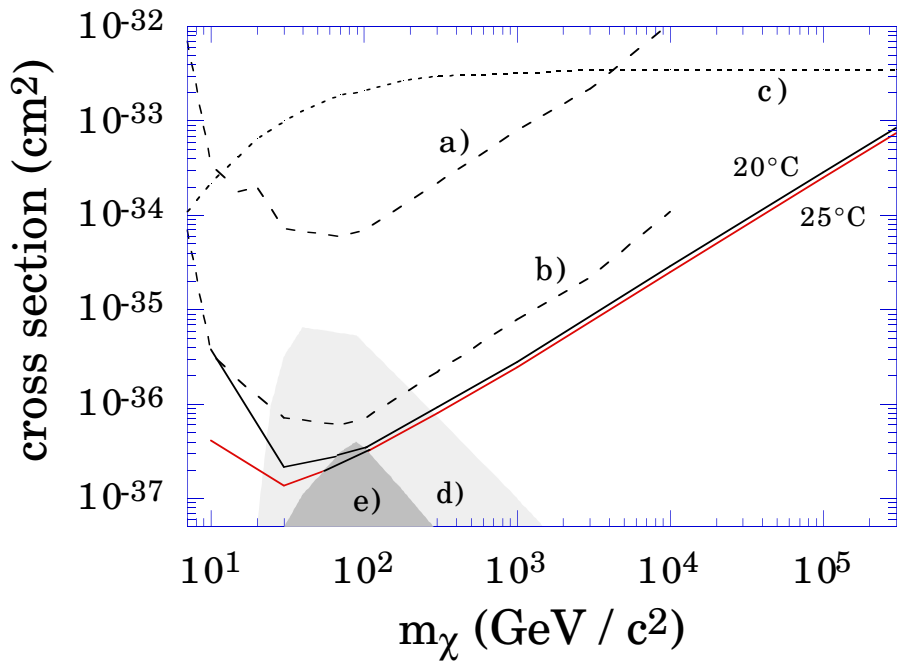

15